# The Impact of AI on Academic Research and Publishing

Brady Lund[1], Manika Lamba[2], Sang Hoo Oh[2]

[1]University of North Texas        [2]University of Illinois at Urbana-Champaign

**Abstract**

Generative artificial intelligence (AI) technologies like ChatGPT, have significantly impacted academic writing and publishing through their ability to generate content at levels comparable to or surpassing human writers. Through a review of recent interdisciplinary literature, this paper examines ethical considerations surrounding the integration of AI into academia, focusing on the potential for this technology to be used for scholarly misconduct and necessary oversight when using it for writing, editing, and reviewing of scholarly papers. The findings highlight the need for collaborative approaches to AI usage among publishers, editors, reviewers, and authors to ensure that this technology is used ethically and productively.

# Introduction

Generative artificial intelligence technologies have rapidly transformed our daily lives, with one of the most profound impacts observed in the realm of writing. These models can produce content at a level that either matches or surpasses the quality of an average human writer. This transformation holds particular significance in academia, where faculty members are traditionally expected to engage in extensive scholarly writing. The increasing prevalence of generative artificial intelligence in academia raises substantial ethical concerns. Reports of scholarly misconduct, spanning a spectrum of issues, have surged in recent years. The implications of integrating such technology into the academic landscape prompt a careful examination of the ethical considerations associated with its use.

The impact of AI on academic research and publishing is multifaceted. Shevlane (2019) highlights the potential for both misuse and protection in AI research, emphasizing the need for a balanced approach. Upshall (2019) discusses the application of AI in scholarly publishing, particularly in identifying suitable peer reviewers. Cheng (2012) provides a quantitative analysis of recent AI research publications, identifying influential journals, languages, and authors. Thelwall (2019) raises concerns about the potential for bias in the peer review process due to AI automation, particularly favoring authors from high-impact countries. These studies collectively underscore the need for careful consideration of the ethical and practical implications of AI in academic research and publishing.

**Definitions of Terms**

- *Generative Artificial Intelligence* (Gen AI, GAI, also called "generative models" in this chapter) is a type of artificial intelligence that generates new data/content rather than only classifying or predicting outcomes based on input data (Lund & Wang, 2023).
- *Machine Learning* (ML) is the process of predicting features of past/trained data on new/test data (Lamba and Madhusudhan, 2022).
- *Large Language Models* (LLMs, also called "language models" and "AI models" in this chapter) "refer to transformer language models that contain hundreds of billions (or more) of parameters, which are trained on massive text data, such as GPT-3, PaLM, Galactica, and LLaMA" (Zhao et al., 2023).

- *Generative Pre-Trained Transformer* (GPT, also referred to when referencing the technology ChatGPT) is the underlying large language model behind the popular platform, ChatGPT. It is characterized by its pattern of training, which consists of unsupervised learning followed by supervised fine-tuning (Alt et al., 2019).

**Selecting a Model**

If using a generative model for academic purposes, it is important to determine which one is the best for your needs. ChatGPT stands out for its ubiquity, but it is a general model. Many models have been trained using training data related to specific topics or purposes (Table 1).

**Table 1: Selected LLM Models (adapted from Zhao et al., 2023)**

| Model (size) | Release Time | Data Source | Availability | Paper |
|---|---|---|---|---|
| T5 (11B) | Oct 2019 | Webpages (100%) | Publicly Available | Raffel et al. (2020) |
| LLaMA (65B) | Feb 2023 | Webpages (87%), Books & News (5%), Code (5%), Scientific Data (3%), Conversational Data (2%) | Publicly Available | Touvron et al. (2023) |
| GPT-3 (175B) | May 2020 | Webpages (84%), Books & News (16%) | Closed Source | Brown et al. (2020) |
| GoPHER (280B) | | Webpages (60%), Books & News (37%), Code (3%) | Closed Source | Rae et al. (2021) |
| GLaM (1200B) | Dec 2021 | Webpages (48%), Conversational Data (30%), Books & News (22%) | Closed Source | Du et al. (2021) |
| LaMDA (137B) | Jan 2022 | Conversation Data (50%), Webpages (38%), Code (13%) | Closed Source | Thoppilan et al. (2022) |
| Galactica (120B) | Nov 2022 | Scientific Data (86%), Webpages (8%), Code (7%) | Publicly Available | Taylor et al. (2022) |
| GPT-NeoX (20B) | Apr 2022 | Scientific Data (38%), Webpages (30%), Books & News (15%), Conversational Data (10%), Code (8%) | Publicly Available | Black et al. (2022) |

| CodeGen (16B) | Mar 2022 | Code (39%), Scientific Data (25%), Webpages (20%), Books & News (10%), Conversational Data (6%) | Publicly Available | Nijkamp et al. (2023) |
| AlphaCode (41B) | Feb 2022 | Code (100%) | Closed Source | Li et al. (2022) |

With these options noted, it is still possible that one would like to use ChatGPT. Its visibility is important. With questions of policy or the law, ChatGPT will be the most scrutinized model. Users should have a good idea of where the model stands in terms of these issues. That may not be the case with other, lesser-used models.

## Ethics of AI for Writing Papers

**Idea Generation**

AI models are adept at idea generation. The better the prompts that you can provide, the better the ideas that you can receive. Yet, it is important to use these AI-generated ideas with care. Considering these models reference their training data, the ideas may not be entirely original. Thus, the models should be used in conjunction with other information sources. If the generation of ideas also involves the inclusion of scholarly references, authors must also be aware of "hallucinations," or fake references created by the model (Sanchez-Ramos et al., 2023).

A good prompt for idea generation will provide specific details. A prompt like "generate topic ideas for a paper relating to developmental psychology" is so broad that it is unlikely to produce particularly meaningful or original ideas. This prompt can be narrowed significantly to read something like "generate topic ideas for a research paper relating to the symptomatology of autism spectrum disorder among Hispanic young adults." This prompt is specific and should be rewarded with specific, new ideas from the AI model.

Another means for generating research ideas would be, instead of focusing on a topic, to provide information about your past research, such as titles and abstracts, and have the model use this information to generate ideas for related studies. This could be a particularly fruitful approach, as it leverages your own past ideas and interests, rather than ideas that may have initially come from another source. Of course, a concern is relevancy and feasibility. A model may present an idea that is unrealistic and it is the role of the researcher to determine which ideas are actually worthwhile. An AI model is a brainstorming tool, not an excuse to go on "auto-pilot" in terms of critical evaluation of ideas.

**Creation of Sample Datasets**

When developing new algorithms or approaches to extracting knowledge from data, a challenge can be finding appropriate datasets with which to test them. This is a challenge that AI models are particularly adept at solving. By providing information about the approach you have created and any data requirements, an AI model can produce a dataset that will work for testing. It can even generate various datasets designed to produce vastly different results or flawed datasets designed to produce an error message. This development is groundbreaking for data scientists and researchers. AI models will work especially well for creating sample datasets when they are provided with a few examples.

**Data Analysis**

Large language models can be used to support both quantitative and qualitative analysis, but this should be done with care. Machine learning models excel with statistical analysis. However, the amount of detail in most research datasets makes it difficult to analyze using web-based models like ChatGPT. Feeding only partial data into a model will significantly impact most analyses a researcher would want to perform. These models, however, can be quite adept at producing Python or R scripts that can then be used to analyze data. For instance, Figure 1 shows a ChatGPT-generated Python script for performing a logistical regression analysis on a dataset with one dependent variable and five explanatory variables. The script, when run in Python, works perfectly.

```
pip install statsmodels

import pandas as pd
import statsmodels.api as sm

# Load your dataset
# Replace 'your_dataset.csv' with the actual file path
df = pd.read_csv('your_dataset.csv')

# Define the dependent variable (response) and explanatory variables (features)
dependent_variable = 'dependent_variable_column_name'
explanatory_variables = ['explanatory_var1', 'explanatory_var2',
'explanatory_var3', 'explanatory_var4', 'explanatory_var5']

# Add a constant term to the explanatory variables matrix
X = sm.add_constant(df[explanatory_variables])

# Define the dependent variable
y = df[dependent_variable]

# Fit the logistic regression model
model = sm.Logit(y, X)
result = model.fit()

# Display the summary of the regression analysis
print(result.summary())
```

**Figure 1. ChatGPT-Generated Python Script for Logistical Regression**

Large language models that do not have the same character restrictions as web-based models like ChatGPT may be able to perform the analyses directly. Additionally, these models can clean and transform data to perform analysis. They can even recommend the best statistical analyses to perform, based on the type and amount of data collected. Many of the popular models can provide a data table and distinguish categorical, nominal, and scalar data.

Large language models can also assist with qualitative analysis. These models have been designed to "understand" human language through computational methods, so tasks like content analysis can be performed quite easily. This is more or less the model's main purpose. However, the issue with qualitative analysis is quality, consistency, and transparency (Jiang et al., 2021). When human researchers perform content analysis, they can agree upon a set of codes and describe a common process of identifying these codes in the data. As of yet, this is not a strength of language models.

So how could LLMs be used to support qualitative analysis, if a researcher was interested in using them for this purpose? They could be used to help identify an initial set of codes that can then be compared to the work of human coders. They could also be used to identify general themes that could then be explored further by human researchers. This could also be helpful for guiding systematic or scoping literature reviews. For instance, it could be used to determine section headings or classify papers underneath existing headings.

**Writing Manuscripts**

Large language models can help to organize a literature review. If you collect your own summaries of publications to be cited, the language model can revise and edit the summaries into a coherent narrative. Conversely, it can determine subheadings for a literature review and identify which articles could fit under each one. This does not replace the work of reading and summarizing papers for a literature review, but it can help to ensure quality of the reviews and perhaps save some time. Additionally, LLMs can compose analyses for the results section of a paper quite well based just on the tables that you create to display findings. However, LLMs are not particularly good at writing entire manuscripts from scratch and submitting papers written by LLMs as your own creates significant ethical questions and issues.

**Editing Manuscripts**

Perhaps the task for which large language models are best suited within academia is revising and editing existing content within manuscripts. By providing a simple prompt like "please revise the following content to improve its quality and clarity," a model like ChatGPT can rework lengthy sections of a paper to read at a level virtually indistinguishable from a professional writer or editor. This is a potentially democratizing power. Authors for whom English is not their primary language now have a tool with which they can work to improve their scholarly writing to a level expected by higher-level academic journals.

However, there are considerable pitfalls that authors must avoid when using LLMs for this purpose. These models are by no means perfect and can produce less than satisfactory outputs on occasion, due to changing the meaning of sentences or inappropriately replacing key terms. Thus, it is critical to check the model's output before replacing any of the text in the original paper. The query may also need to be run through the model multiple times before a suitable output is procured.

## AI Policies Among Publishers

*AI Authorship Attribution.* The matter of authorship attribution with artificial intelligence is a challenging one. In the early months following the release of ChatGPT, many writers included the language model as an additional author in their publications (King & ChatGPT, 2023). However, by late spring, many journal publishers had posted a policy on their sites indicating that the use of language models should be acknowledged within the paper but these models should not be listed as authors (Lund & Naheem, 2023). The rationale for not including AI as an author is due to the common criteria used to determine authorship for scholarly articles: that the author must contribute to the writing of the article, that the author must be able to understand all that is written, and that the author must be able to take responsibility for what is written. The final two points are a barrier to including AI as an author, as it is not clear that an AI model can understand, consent, and take responsibility for the content that it produces (Titus, 2024).

Additionally, existing AI authorship policies note that the AI models should be used for improving the quality of written works, rather than to generate new content from scratch. When asked to generate new content, language models rely on their training data, which means generated content is not necessarily original and is subject to copyright infringement.

There is some debate as to whether attribution of a large language model is actually necessary if that model is simply used as a tool to enhance writing and grammar, rather than to generate altogether new content. Suppose the model is not assimilating new knowledge into the existing writing, per se. In that case, its role is not significantly different from the use of tools embedded into Microsoft Word, like spell and grammar check, to improve writing. However, given the "black box" nature of AI, where users cannot tell how the tool actually works, concerns about the technology persist. The policy of acknowledging AI use may be a stop-gap measure until the nature of these models can be better understood.

*AI Policies.* Ultimately, the extent to which AI usage is allowed or barred is likely to remain at the discretion of individual publishers. Like with individual people, some publishing companies may be eager adopters and allow and integrate AI into many aspects of the academic publishing process.

One thing that seems clear is that each publisher needs to provide a clear policy on whether AI is allowed and, if so, in which aspects of manuscript preparation and review. Without these policies in place, it is a veritable "wild west" for authors, reviewers, and editors (Lund et al., 2023). The following is an example of a policy a publisher might use:

> The use of artificial intelligence tools, including large language models like ChatGPT, is permitted only for the purposes of enhancing the quality of writing in a completed manuscript. Authors are encouraged to retain a copy of their paper before AI was sued to make revisions and should be prepared to supply this copy to the editors if any questions about the originality of a manuscript arises. Peer reviewers for this journal may not use AI tools when preparing their reviews. Authors who utilize AI should also acknowledge its usage (including the model and date(s) used) in the acknowledgements of their paper.

The above policy outlines how authors and reviewers may (or may not) use AI tools. This would likely limit some negative press should it be found that some authors have used AI without acknowledgement. Additionally, it gives authors clarity about journal practices before they submit.

Whatever the policy, editors should practice caution with AI, in order to protect themselves and their publications. If lawsuits surrounding the use of copyrighted materials to train AI progress, then the products of that model may also be challenged (Grynbaum & Mac, 2023). By having a policy that disallows the use of AI to write substantial paper content, publishers can hopefully mitigate this potential issue.

## AI in Editorial Processes

### AI for Determining Value of Manuscripts

Could AI be used to provide a preliminary review of a paper to determine if it fits the scope and quality requirements for an academic journal?

Previous research has shown that different AI tools can be used to provide a preliminary review of submitted manuscripts (Checco et al, 2021; Kousha & Thelwall, 2023). With the advent of large language models, AI manuscript management tools are now able to automate initial quality control of submitted manuscripts to determine if they fit the scope and quality requirements of an academic journal. Initial quality control-related tasks that these AI manuscript management tools can perform include the following: plagiarism detection, robot author detection, methods and automated statistical checking, transparency and reproducibility checking, manuscript structure checking, and multipurpose manuscript evaluation (Kousha & Thelwall, 2023). It is evident that AI manuscript management tools can go beyond basic editorial tasks such as plagiarism and robot author detection. They are now capable of performing more complex tasks such as statistical checking and multipurpose manuscript evaluation. While the accuracy of these AI manuscript management tools needs to be further examined, it is clear that they hold significant promise for the future application of AI in the preliminary review process of a journal paper.

**AI for Peer Review**

There is evidence that large language models have already been used in some cases to either supplement or replace the duties of peer reviewers. Anecdotally, among the authors of this chapter, evidence of an AI-generated peer review for one of their papers was found on five occasions in 2023 alone. Some aspects of AI-generated reviews can be useful. For instance, suggestions to improve the writing and grammar of a paper can be useful. However, suggestions related to the actual content of a paper are often severely lacking. Most LLMs are trained in large stores of general knowledge but are not experts in specific subject areas. Existing language models lack the judgment of the human expert whose feedback has been requested by a journal or conference. One day, such models may exist - models that are trained to be experts in a specific subject area and assess the quality of new contributions - but a model like ChatGPT does not satisfy this objective.

LLMs could be used to help organize peer reviews and better articulate the reviewers' concerns. They could make suggestions as to the structure and grammar of the paper. They could also be used for the initial screening of papers to ensure relevance to the journal and for the identification of relevant peer reviewers (Checco et al., 2021; Kousha & Thelwall, 2023). However, these models should not be used to complete an entire review. The journal/conference has requested a review from a human reviewer, based on that individual's expertise, not ChatGPT. Furthermore, if the journal wanted a review from ChatGPT, they could simply ask ChatGPT directly, rather than ask a reviewer to use the model to generate a review for them.

**AI for Editing Tasks**

AI-assisted decision support systems for editors, reviewers, and authors may aid in decision-making and speed up the overall process (Ghosal, 2019). Further, AI-based manuscript writing support may allow the researcher to perform creative work in a more refined fashion (Nakazawa, 2022).

As opposed to peer reviewers using large language models to support peer review, it makes considerably more sense for journal or proceedings editors to utilize it in order to suggest potential edits to improve the quality of a manuscript. This is the flip side of encouraging authors to use AI directly to improve their papers. If editors do not believe that authors should be given the authority to use AI, but believe that they can use it more judiciously to make suggestions for the authors, then this may be an approach that they could take. Still, some publishers will likely be opposed to any use of AI at all, considering it to be unfair to talented writers who do not require AI assistance.

## Conclusion

AI has already dramatically transformed academia and scholarly publishing in recent years, with the launch of ChatGPT representing an acceleration of this trend. It is critical for publishers to consider the entire research and publishing process when developing policy. A complete ban on AI use is likely not prudent, given a society that is increasingly adopting it in order to gain a competitive edge, but some restrictions on how AI can be used are necessary. Likely, a collaboration between publishers and their editors and authors would produce the most fruitful results.

As for researchers, it is advisable to use AI with care. These AI models are new and there are many ethical and legal issues yet to be sorted out. Recent years have shown a spike in article retractions, which can ruin promising academic careers. Avoidance of AI usage unless explicitly permitted in a publisher's policy is likely the best approach. Undoubtedly, more issues will emerge as AI technology matures and becomes more sophisticated. Authors, editors, and publishers have an opportunity to lead discussions on these issues and communicate clearly with one another in order to mitigate disruption.